# To Act or not to Act? Political competition in the presence of a threat

Arthur Fishman[a,b] and Doron Klunover[a*]

November 2020

**Abstract**

We present a model of political competition in which an incumbent politician may implement a costly policy to prevent a possible threat to, for example, national security or a natural disaster. A competent politician is privately informed about the actual probability of disaster while an incompetent one only knows the average probability. After the policy is implemented, it may be unknown whether or not it was required. We show that, in these circumstances, a competent politician only takes advantage of her private information about the likelihood of disaster when voters are imperfectly rational, and vote only on the basis of the policy's past outcome and not on what it may reveal about the incumbent's future performance. Our result is consistent with and may help explain voters' behavior observed after natural disasters.

[a] Department of Economics, Ariel University, 40700 Ariel, Israel.
[b] Department of Economics, Bar-Ilan University, 52900 Ramat-Gan, Israel .
[*] Corresponding author: E-mail address: doronkl@ariel.ac.il

1. **Introduction**

We consider an incumbent politician whose country faces the threat of a possible natural disaster. If no action is taken, the disaster will occur with some imperfectly known probability. Alternatively, the incumbent may take appropriate measures to avert the disaster. But both action and inaction pose risks for her prospect of reelection. She will not be reelected if disaster strikes. But voters may also punish her if actions are taken which voters subsequently judge to have been unnecessary and wasteful.

As an example, consider the California wildfires of 2019. After widespread outcry, the public demanded to know why controlled burns were insufficiently used as a preventative measure. Miller et al. (2020) studied this issue and found that lack of aggressive controlled burns in the fire-plagued state was related to costs of labor and equipment and the potential for property damage. Now, greater consensus has emerged that adopting the more aggressive (and costly) controlled burns approach is required. However, this leaves open the potential future criticism that if wildfires do not re-occur such a policy may be judged to be overly aggressive and unnecessarily expensive.

We develop a model to study this issue when the incumbent may be competent – meaning that she is better informed (or is able to take steps to become better informed) than the average voter about the appropriate policy choice, or incompetent – meaning that she is no better informed than the average voter. The incumbent is privately informed of her type but voters are not. We analyze equilibrium outcomes for two types of voters. Backward looking voters vote only on the basis of the chosen policy's outcome without considering what it may reveal about the incumbent's competence – which is relevant for her response to future events. Forward looking voters, by contrast, are more rational and also consider what the chosen policy reveals about the incumbent's level of competence. We show that in equilibrium, the incumbent adopts a better policy - i.e. one which is more aligned with voters' interests - when voters are backward looking than when they are forward looking. In particular, when voters are backward looking, the competent incumbent's choice of policy is guided by her superior private information. When voters are forward looking, by contrast, the competent incumbent's policy ignores her private information. Thus our analysis suggests that backward looking behavior of voters following natural disasters (e.g. floods, wildfires, virus outbreak etc.) may lead to better policy choices.

Voter and incumbent behavior in the context of natural disasters has been studied empirically. In particular, Healy and Malhotra (2009) showed that voters reward the incumbent presidential party for delivering disaster relief spending, but not for investing in disaster preparedness spending. Gasper and Reeves (2011) showed that voters punish politicians after severe weather damage. Achen and Bartels (2004) found that voters regularly punish governments for droughts, floods, and shark attacks. Malhotra and Kuo (2008) designed a survey experiment to discover which public official citizens blamed after Hurricane Katarina and concluded that although voters are not objective processors of information as envisioned by Kramer (1971), they are also not myopic. Wolfers (2002), who considered voters to be rational only if they distinguish between lucky and competent incumbents, showed that voters vote for incumbents after national booms, dumping them after national recessions. Those studies suggest that voter behavior tends to be backward looking and our analysis suggests that when facing a collective threat such backward looking voter behavior may be beneficial.

The rest of the paper is structured as follows: In the remainder of this section, we present a survey of the literature. In section 2, we describe the model. Sections 3 characterizes equilibria when voters are backward looking and section 4 characterizes equilibria when they are forward looking. Welfare consequences are analyzed in section 5. In section 6 we discuss voters with different preferences than the ones assumed in the model. Section 7 concludes.

**Related Literature**

Our analysis is closely related to the literature about the effect of imperfect voter rationality on politicians' behavior. Several studies have shown that Bayesian failures or cognitive biases of voters can be beneficial. For instance, Ashworth and Bueno de Mesquita (2014) shows that behavioral biases might be beneficial for voters when considering strategic politicians, Levy and Razin (2015) demonstrate that voters who have a "correlation neglect" cognitive bias improves political outcomes, while Millner et al. (2020) show that it can be beneficial when voters have "confirmation bias". By contrast, in our model voters always "correctly" (using Bayes' rule) process all available information. Our analysis therefore can be viewed as complementary to these papers by showing that backward looking Bayesian voters may lead to better policy outcomes.

We are also related to the herding literature, in which decision makers ignore their private information. The most closely related paper in this vein is Scharfstein and Stein (1990). They consider a model with two reputation-oriented ex-ante identical managers, each of whom who may be "smart" or "dumb". Each manager receives a private informative signal about the profitability of an investment opportunity and make sequential investment choices. They show that when smart managers' signals are more correlated than those of dumb managers, the second investor ignores her private signal and mimics the first investor's behavior, who, in contrast, always follows her own private signal. The reason is that as smart investors' signals are more correlated than those of dumb investors, mimicking the first investor implies that they received identical signals, which increases the probability that the second investor is smart. Thus in their model, a single decision maker (or the first one) always acts based on her private information. By contrast, in our model there is a single decision maker who ignores her private signal in equilibrium.[1]

---

[1] Another important difference is that Scharfstein and Stein (1990) assume that, ex-poste there is complete information on the state of the world, while ex-poste incomplete information on the state of the world is a key property of our model.

Our research is also related to the literature on government accountability that studies how the desire of politicians to be reelected affects policy choices (see, Maskin and Tirole, 2004; Battaglini and Harstad, 2020; Herrera et al. 2020), and to the literature on politician quality that addresses two questions: 1) Who becomes a politician? (Caselli and Morelli, 2004; Dal Bo et al., 2017) and 2) Where are the best politicians located? (Galasso and Nannicini, 2011). In particular, Caselli and Morelli (2004) argue that less qualified citizens become politicians due to comparative advantage. Our result suggests that qualified politicians choose better policies when voters are willing to vote for unqualified ones.[2]

## 2. The model

An incumbent politician faces the threat of a natural disaster to her constituency. Specifically, there are two possible states of the word, denoted $S$. If $S=B$ (bad) the disaster will strike unless the incumbent acts by taking appropriate, and costly, measures to prevent it. If $S=G$ the disaster will not occur. $S=B$ with probability $p'$ and $S=G$ with probability $1-p'$, where $p'$ is a random variable with a commonly known probability distribution $f$ over the interval $[0,1]$ and mean $p>0$. The incumbent chooses between two policies: either to *act* – taking costly measures to prevent the disaster – or *not to act*. We denote the incumbent's policy choice by *c*, where *c=a* if she acts and *c=na* if she doesn't. The incumbent must choose her policy before the actual state is possibly revealed as described below.

---

[2] Cohen and Werker (2008) considered a model of policy choice in the context of natural disasters, in which there is a common prior on the state of the world and the government is viewed to be a social planner. They show that international aid increases the chance that governments will under-invest in natural disaster policy. Note that, in contrast to their model, private information is the key property of our model. In particular, we assume that the incumbent has private information on the state of the world if she is competent and whether or not she is competent is private information as well.

The incumbent may be either competent or incompetent. A competent incumbent knows the actual $p'$. An incompetent incumbent only knows the average probability $p$.[3] The incumbent is privately informed of her type but voters are not; they assign prior probability $q$ that the incumbent is competent. Each type of incumbent chooses her policy to maximize the probability of being reelected. At the end of the incumbent's tenure a single voter decides whether to reelect the incumbent or to elect the challenger.

We assume the voter's utility is highest if there is no disaster and no action is taken, is reduced by the cost of action if action is taken, and is lowest if the disaster occurs. In addition we assume that the cost of *a* is *p* – that is, the cost of prevention is proportional to its likelihood. Then the voter has the following utility function:

(1) $$u = \begin{cases} 1-p, & \text{if } c = a \\ 0, & \text{if } c = na \text{ and } S = B \\ 1, & \text{if } c = na \text{ and } S = G \end{cases}.$$

---

[3] A competent incumbent is able to interpret a private signal revealed to her at the beginning of her tenure, while an incompetent incumbent is unable to.

The voter's most preferred policy is the one that maximizes her expected utility. (1) implies that voter's expected utility is maximized when $c=a$ if $p'>p$, and is maximized when $c=na$ if $p'<p$. In other words the voter considers the expected benefits from $a$ to exceed the cost when the probability of disaster is above average and considers the cost to exceed the expected benefit when it is below average. Given the high level of uncertainty involved in natural disasters, this assumption seems to be reasonable, at least in some circumstances. We refer to a voter with these preferences as *action neutral*.[4]

At the end of the incumbent's tenure, the voter observes the policy choice and whether or not the disaster has taken place. In addition, with exogenous probability $1-y$, $y\epsilon[0,1]$ the true state $S$ is publicly revealed before voting takes place if the incumbent has chosen $c=a$ (and with probability $y$ is not revealed). For instance, with a probability $y$ a floodgate completely blocks a flood and, therefore, ex-post it will be impossible to know whether or not a flood would have occurred. However, with probability $1-y$, the flood is partly blocked such that some harmless amount of water passes the floodgate, providing evidence that the floodgate successfully prevented flooding.

Thus, if $c=na$, then $S$ is always revealed (since when $c=na$ disaster is averted only if $S=G$). But if $c=a$, then $S$ is only perfectly observed with probability $1-y$. Finally, the voter decides whether to vote for the incumbent or a challenger.

---

[4] This may be viewed as the preferences of the median voter, while other voters are either *pro-action* or *against-action*. In section 6 we consider voters with such preferences.

The sequence of events is shown below.

**Timeline**

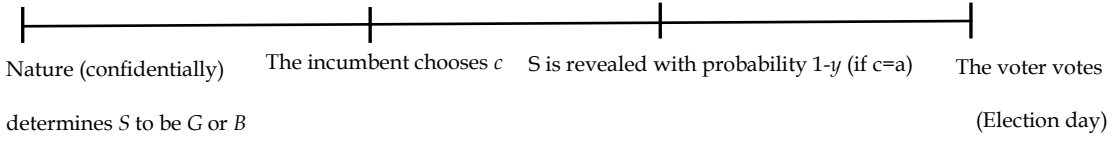

Nature (confidentially) determines S to be G or B    The incumbent chooses $c$    S is revealed with probability 1-$y$ (if c=a)    The voter votes (Election day)

### 3. Backward looking voters

In this section we assume that the incumbent is reelected only if, given the available information, her chosen policy is perceived to have maximized voter's expected utility on election day, irrespective of what the policy choice might reveal about the incumbent's competency. We refer to this behavior as *backward looking*.

Formally, let $p^{el}$ be the probability, as perceived by the voter on election day, that S=B. Then, $p^{el}$=0 if S=G is revealed, $p^{el}$=1 if S=B is revealed, and $p^{el}$ is given by Bayes' rule if $S$ is not revealed. Let $r_{bl}$ be the probability that a backward looking voter votes for the incumbent after the latter is observed to choose $c\epsilon\{a,na\}$. Then by (1),

$$(2)\ r_{bl} = \begin{cases} 1, \text{if } c = a \text{ and } p^{el} > p \text{ or } c = na \text{ and } p^{el} < p \\ 0, \text{if } c = a \text{ and } p^{el} < p \text{ or } c = na \text{ and } p^{el} > p \\ 1/2, \text{if } \hspace{6em} p^{el} = p \end{cases}.$$

*Equilibrium.*

This is a game of asymmetric information since the incumbent is privately informed about her type and the competent incumbent is privately informed about $p'$. Thus an appropriate solution concept is Bayesian equilibrium.

Denote the (possibly mixed) strategies of the competent and incompetent incumbent as $\pi_c(p')$ and $\pi_{nc}$ respectively, where $\pi_c(p')$ is the probability, given $p'$, that a competent incumbent chooses $c=a$ and $\pi_{nc}$ is the probability (given $p$) that the competent incumbent chooses $c=a$.

A Bayesian Equilibrium is a strategy profile $(\pi_c(p'), \pi_{nc})$ such that (i) for each type of incumbent there does not exist a different strategy that increases her probability to be reelected and such that (ii) voters beliefs are consistent with Bayes' rule whenever it applies.

Proposition 1 establishes that the equilibrium policy choice by each type of incumbent is characterized by a threshold probability – she acts if and only if the probability that $S=B$ is above a threshold.

**Proposition 1** *Suppose that the voter is backward looking. Then, the Bayesian equilibria are as follows:*

a. *A competent incumbent chooses a (na) when $p'>(1-y)/(2-y)$ ($p'<(1-y)/(2-y)$).*
b. *An incompetent incumbent chooses a (na) when $p>(1-y)/(2-y)$ ($p<(1-y)/(2-y)$).*
c. *If $p'=(1-y)/(2-y)$ ($p=(1-y)/(2-y)$), then a competent (an incompetent) incumbent's choice is any convex combination of a and na.*[5]

---

[5] Note that in the special case in which $y=1$ and therefore $(1-y)/(2-y)=0$, $\pi_c(0)<1$.

All proofs appear in the appendix. Notice that the equilibrium is unique except for the knife edge cases $p'=(1-y)/(2-y)$ and $p=(1-y)/(2-y)$. Note that the competent incumbent's policy depends on her private information about the actual probability, $p'$, while the incompetent incumbent always acts if $p$ is sufficiently high and never acts otherwise.

This result is very intuitive. The incumbent must balance the risk of not acting – in which case she won't be reelected if $S=B$ - against the risk of acting when actually $S=G$ – in which case she won't be elected if S is revealed. For the competent incumbent, the former risk is greater the higher is $p'$ and for the incompetent incumbent, the risk is greater the higher is $p$. The threshold values of $p'$ and $p$ in the preceding proposition are derived to balance these risks. In the proof of the proposition it is also shown that this threshold strategy ensures that $p^{el}>p$ if $S$ is not revealed and thus when $c=a$ the incumbent is always reelected if $S$ is not revealed.[6]

Two extreme cases are of special interest. First, suppose that $S$ is always revealed (i.e., $y=0$ and therefore $(1-y)/(2-y)=½$). Then, to maximize her chance of reelection, the competent incumbent chooses $a$ when $p'≥½$, and the incompetent incumbent does the same when $p≥½$. At the other extreme, suppose that $S$ is never revealed (i.e., $y=1$ and therefore $(1-y)/(2-y)=0$). Then, both types of incumbents choose *a* except if *p'*=0, in which case the competent incumbent chooses *na* with positive probability.[7]

---

[6]Note that $p^{el}$ is greater that $p$ when $S$ is not revealed iff $\pi_c(p')$ is monotonically increasing in $p'$. See Lemma 1 in the appendix for the detailed proof.

Note that Proposition 1 implies that, by Bayes' rule, the probability that the incumbent is competent conditional on choosing *a* is smaller than *q* when $p>(1-y)/(2-y)$. This is because, in equilibrium, the incompetent incumbent always chooses *a* while the competent one chooses *a* with a probability smaller than 1. Therefore, if the challenger is also competent with probability *q*, an incumbent who chooses *a* is expected to be less competent than her challenger but nevertheless may be reelected. In the next section, we consider a more forward looking voter who will not vote for the incumbent in such a case.

## 4. Forward looking voters

In this section, we consider a voter who cares not only about the outcome of the incumbent's policy but also about what the chosen policy reveals about her competence. More specifically, he does not vote for the incumbent if he believes that she is less competent than the challenger. Such a voter wishes to distinguish an incompetent incumbent who got lucky from one who chose the correct policy out of competence. This is somewhat consistent with the conventional definition of voter rationality which requires that voters filter ability from luck (for instance, see Wolfers, 2002).

---

[7] Since the competent incumbent knows that *S=G* when *p'*=0, and because this strategy ensures that $p^{el}>p$ when *S* is not revealed (which is always the case when *c=a*), she is willing to mix between *a* and *na* when *p'*=0. The incumbent is therefore always elected when *S* is never revealed.

Formally, we define forward looking voters as follows. Let $q^c$ be the probability that the incumbent is competent conditional on choosing $c \epsilon \{a,na\}$ and let $r_{fl}$ be the probability that a forward looking voter votes for the incumbent. Then: $r_{fl}=0$ if $q^c<q$. If $q^c \geq q$, then $r_{fl}=r_{bl}$, where $r_{bl}$ is given by (2).

The following proposition presents equilibria when voters are forward looking.

***Proposition 2*** *When voters are forward looking, there are exactly two Bayesian equilibria: either $(\pi_c,\pi_{nc})=(0,0)$ for all $p'$ or $(\pi_c,\pi_{nc})=(1,1)$ for all $p'$.*

Thus, in either equilibrium, the competent incumbent's strategy is independent of $p'$ when the voter is forward looking and, therefore, the incumbent's policy choice is not informative.

The reason for this result is as follows. Suppose that the incumbent plays a pure strategy and that, to the contrary, the competent incumbent's choice depends on $p'$. Then, since the incompetent incumbent is uninformed about $p'$, she either always chooses $c=a$ or always chooses $c=na$. Suppose she chooses $c=a$. Then, voters who observe $c=a$ will update beliefs to $q^c<q$. In that case an incumbent who chooses $a$ is never elected and therefore neither type of incumbent will ever choose $a$. A parallel argument applies if the incompetent incumbent always chooses $na$. An extension of this reasoning to show that there is no equilibrium in which the incumbent plays a mixed strategy is somewhat more involved and appears in the appendix.

The next section considers welfare consequences of the voter's behavior.

## 5. Voter Welfare

The purpose of this section is to compare the effect of the equilibrium policies on the utility of the backward looking and forward looking voter.

Let $U_b$ be the backward looking voter's expected equilibrium utility, let $u_c$ be his expected equilibrium utility if the incumbent is competent and $u_{nc}$ his expected equilibrium utility if the incumbent is incompetent. Then

$U_b = qu_c + (1-q)u_{nc}$.

By Proposition 1,

$$(3)\ u_c = \int_{p'=0}^{\frac{1-y}{2-y}} f(p')(1-p')dp' + \int_{p'=\frac{1-y}{2-y}}^{1} f(p')(1-p)dp'$$

$$= 1 - \left( \int_{p'=0}^{\frac{1-y}{2-y}} f(p')p'\, dp' + \int_{p'=\frac{1-y}{2-y}}^{1} f(p')p\, dp' \right).$$

By proposition 1, depending on the value of $p$, the incompetent incumbent either always chooses $c=a$ or always chooses $c=na$. By (1), in either case the voter's expected utility is $1-p$. Thus $u_{nc}=1-p$ and thus:

$U_b = qu_c + (1-q)(1-p)$.

Let $U_f$ be the expected equilibrium utility of a forward looking voter. By proposition 2, when the voter is forward looking both types of incumbent either always choose *a* or always choose *na*, and again, by (1), the voter's expected utility is *1-p* in either case. Thus, $U_f$=*1-p* and thus:

$U_b - U_f = q(u_c - (1-p))$ where $u_c$ is given by (3).

***Proposition 3:*** $U_b - U_f > 0$

Proposition 3 therefore establishes that backward looking voter's equilibrium expected utility is higher than that of the forward looking voter.

## 6. Voter with different preferences

The preceding analysis has assumed that the voter is ex-ante action neutral. In this section we examine the extent to which our results generalize when this assumption is relaxed. In particular, the voter utility function is now:

$$(1') \quad u = \begin{cases} p/\underline{p} - p, & \text{if } c = a \\ 0, & \text{if } c = na \text{ and } S = B \\ p/\underline{p}, & \text{if } c = na \text{ and } S = G \end{cases},$$

where a voter is considered to be *pro-action* if $\underline{p}$<*p*, *against action* if $\underline{p}$>*p*, and *action neutral* if $\underline{p}$=*p* (in which case (1') corresponds to (1)).

***Proposition 4***

(i) *If the voter is backward looking and pro action, then the incumbent's equilibrium policies are given by Proposition 1*

(ii) *If the voter is forward looking, then the incumbent's equilibrium policy is given by proposition 2 whether the voter is pro action or against action.*

Thus the equilibria are robust to changes in voter preferences when the voter is forward looking and when the voter is backward looking and pro action. However, this is not necessarily the case when the voter is backward looking and against action. Recall that one reason the strategy described in Proposition 1 is optimal for the incumbent is that it ensures that $p^{el}>p$ and her reelection when $S$ is not revealed. Thus if the voter is pro action (i.e., $p>\underline{p}$) the strategy also implies a fortiori that $p^{el}>\underline{p}$ when $S$ is not revealed. By contrast, if the voter is against action, (i.e., $p<\underline{p}$) it may be that $p^{el}<\underline{p}$ when $S$ is not revealed and hence the threshold policy described in proposition 1 might no longer be optimal. The equilibrium policy will then depend on model parameters. We discuss this case more fully in Appendix 2.

Although the equilibrium policies when voters are pro action are unchanged, the welfare consequences of those policies may differ from those derived above in section 5. Recall that when the voter is *action neutral* (is indifferent between the policies when $p'=p$), then the incompetent incumbent's policy has the same effect on utility of backward and forward looking voters. And similarly, when the voter is action neutral each equilibrium policy given by proposition 2 has the same effect on the utility of a forward looking voter. By contrast, when voters are not action neutral, they strictly prefer one policy over the other when $p'=p$, and thus the policy choice has welfare consequences both when the voter is backward and when she is forward looking. As a result, welfare comparison between backward and forward looking voters depends on $p$ (as given by the threshold strategy in proposition 1) when voters are backward looking and the equilibrium selection when they are forward looking.[8]

## 7. Conclusion

We consider a model of political competition in the presence of a threat. We solve this model when voters are either backward looking or forward looking to show that under certain conditions, competent incumbents' choices are less distortive when voters are backward looking, consistent with widely observed behavior following natural disasters. Our analysis thus suggests that such behavior may be beneficial.

**Appendix 1: Proofs**

Proof of Proposition 1: First, we formally describe the incumbent's problem. Let $r^a(\in\{0,1/2,1\})$ be the probability that the incumbent is reelected when $S$ is not revealed and $E\{r_c|p'\}$ and $Er_{nc}$ the expected probabilities that a competent and incompetent incumbent are reelected, given the information available to them, correspondingly. By (2), a competent incumbent then solves:

(4) $\max_{\pi_c} E\{r_c|p'\}$

where

$E\{r_c|p'\} = \pi_c(p'(1-y) + yr^a) + (1-\pi_c)(1-p')$.

---

[8] In addition to the above mentioned effect of voter's preferences on their expected utility, note that the expected utility of a backward looking voter also depends on his preferences when the incumbent is competent. In particular, since (1') depends on voters' preferences, (3) does as well.

While an incompetent incumbent solves:

(5) $\max\limits_{\pi_{nc}} Er_{nc}$

where

$Er_{nc} = \pi_{nc}(p(1-y) + yr^a) + (1-\pi_{nc})(1-p).$

We now present an auxiliary lemma regrading $p^{el}$.

***Lemma 1*** *If S is not revealed,*

$p^{el} > p$, *if $\pi_c(p')$ is non decreasing in $p'$ and there exists $\hat{p} \in [0,1)$ such that*

$\pi_c(p') > \pi_c(\hat{p})$ *for all $p' > \hat{p}$.*

$p^{el} < p$, *if $\pi_c(p')$ is non increasing in $p'$ and there exists $\hat{p} \in (0,1]$ such that*

$\pi_c(p') > \pi_c(\hat{p})$ *for all $p' < \hat{p}$.*

$p^{el} = p$, *if $\pi_c(p')$ is constant.*

Proof of Lemma 1: When *S* is not revealed, by Bayes' rule:

(6) $p^{el} = \dfrac{q \int_{p'=0}^{1} f(p')p'\pi_c(p')dp' + (1-q)p\pi_{nc}}{q \int_{p'=0}^{1} f(p')\pi_c(p')dp' + (1-q)\pi_{nc}}$

$= \dfrac{q \int_{p'=0}^{1} f(p')p'\pi_c(p')dp' + (1-q)p\pi_{nc}}{qE[\pi_c] + (1-q)\pi_{nc}}.$

Therefore,

$p^{el} \gtreqless p$

$\leftrightarrow$

$$\int_{p'=0}^{1} f(p')p'\pi_c(p')dp' \gtreqless pE[\pi_c]$$

↔

$$E[p'\pi_c] - (Ep')E[\pi_c] \gtreqless 0$$

↔

$$cov(p', \pi_c(p')) \gtreqless 0.$$

The result in Lemma 1 follows by the definition of covariance of two random variables. QED

Note that by Lemma 1 and by (2), when $S$ is not revealed, both $sign(p^{el}-p)$ and $r^a$, are independent of $\pi_{nc}$, which also implies that (4) is independent of $\pi_{nc}$. Therefore, we first find the $\pi_c$ that solves (4), which is independent of $\pi_{nc}$. Then we substitute this $\pi_c$ into (5) and find the $\pi_{nc}$ that solves (5).

Assume that $p^{el}>p$ when $S$ is not revealed. Then, by (2), when $S$ is not revealed, $r^a=1$. Then $E\{r_c|p'\}$ is a convex combination of the two constants: $p'(1-y)+y$ and $1-p'$, and, therefore, is maximized when all the weight is allocated to the larger term, which implies that $\pi_c=0$ ($\pi_c=1$) when ($p'<(1-y)/(2-y)$) ($p'>(1-y)/(2-y)$). And $\pi_c \in [0,1]$ when $p'=(1-y)/(2-y)$. By Lemma 1, when $S$ is not revealed and $\pi_c(p')$ follows this rule, $p^{el}>p$.

Note that if a competent incumbent chooses a different strategy (namely, a different, $\pi_c(p')$) that results in $r^a<1$, then $E\{r_c|p'\}$ strictly decreases for all $p'>(1-y)/(2-y)$ (since of the two terms in (4), the larger one decreases) and does not increase for all $p'≤(1-y)/(2-y)$ (since of the two terms in (4), the larger one does not change). Therefore, the strategy of a competent incumbent, $\pi_c(p')$, is uniquely determined by the above rule.

Substituting $r^a=1$ in (5) gives $Er_{nc} = \pi_{nc}(p(1-y)+y) + (1-\pi_{nc})(1-p)$, which implies that the optimal strategy of an incompetent incumbent is $\pi_{nc}=0$ ($\pi_{nc}=1$) when $p<(1-y)/(2-y)$ ($p>(1-y)/(2-y)$). And $\pi_{nc}\epsilon[0,1]$ when $p=(1-y)/(2-y)$.     QED

Proof of Proposition 2: We first present an auxiliary lemma, which reduces the possible equilibria strategies to two, and then proceed to prove the proposition. Note that below we use notations defined in the beginning of the proof of Proposition 1.

**Lemma 2** *If the voter is forward looking, then in a Bayesian equilibrium, either $(\pi_c,\pi_{nc})=(0,0)$ for all p' or $(\pi_c,\pi_{nc})=(1,1)$ for all p'.*

Proof: Note that in equilibrium, $\pi_{nc}\epsilon\{0,1\}$. This can be proved by contradiction. Assume that $\pi_{nc}\epsilon(0,1)$. This is only possible if $q^c=q$ for all $c\epsilon\{a,na\}$; otherwise either $q^a<q$ which implies that $\{r_{fl}|c=a\}=0$ and therefore the incumbent will deviate to $\pi_{nc}=0$, or $q^{na}<q$ which implies that $\{r_{fl}|c=na\}=0$ and therefore the incumbent will deviates to $\pi_{nc}=1$.[9] Now, assume that, $\pi_{nc}\epsilon(0,1)$ and that $q^c=q$ for all $c$. By (5), an incompetent incumbent's best response is to choose

---

[9] Note that $(q\int_{p'=0}^{1} f(p')\pi_c(p')dp'+(1-q)\pi_{nc})q^a+(q\int_{p'=0}^{1}f(p')(1-\pi_c(p'))dp'+(1-q)(1-\pi_{nc}))q^{na}$
$=(q\int_{p'=0}^{1}f(p')\pi_c(p')dp'+(1-q)\pi_{nc})\frac{q\int_{p'=0}^{1}f(p')\pi_c(p')dp'}{q\int_{p'=0}^{1}f(p')\pi_c(p')dp'+(1-q)\pi_{nc}}+(q\int_{p'=0}^{1}f(p')(1-\pi_c(p'))dp'+(1-q)(1-\pi_{nc}))\frac{q\int_{p'=0}^{1}f(p')(1-\pi_c(p'))dp'}{+(q\int_{p'=0}^{1}f(p')(1-\pi_c(p'))dp'+(1-q)(1-\pi_{nc}))}=q$. Therefore, since $q\int_{p'=0}^{1}f(p')\pi_c(p')dp'+(1-q)\pi_{nc}+q\int_{p'=0}^{1}f(p')(1-\pi_c(p'))dp'+(1-q)(1-\pi_{nc})=1$, it is impossible that $q^c<q$ for all $c\epsilon\{a,na\}$ or $q^c>q$ for all $c\epsilon\{a,na\}$.

$\pi_{nc}=0$ ($\pi_{nc}=1$) when $p(1-y)+yr^a < 1-p$ ($p(1-y)+yr^a > 1-p$), a contradiction.

In equilibrium if $\pi_{nc}=0$ ($\pi_{nc}=1$), then $\pi_c(p')=0$ ($\pi_c(p')=1$) for all $p'$. This, again, can be shown by contradiction. Assume that the strategy profile is: $(\pi_c(p'),\pi_{nc})=(\pi_c(p'),k)$, where $k\in\{0,1\}$ and there exists $p'\in[0,1]$ for which $\pi_c(p')\neq k$. Then, $q^c<q$ for the specific $c$ not chosen by the incompetent incumbent and therefore the competent incumbent will never choose this specific $c$, a contradiction. QED

The above implies that, a Bayesian equilibrium is therefore a pair: $(\pi,\mu(q^c))$, where $\pi\in\{0,1\}$ is the incumbent's strategy and $\mu(q^c)$ is the voter's beliefs over $q^c$ for the particular $c$ not chosen by the incumbent (i.e., $c=a$ when $\pi=0$ and $c=na$ when $\pi=1$). Let $\mu(q^a)<q$. If $\pi=0$, then: $Er_{nc}>0$ for all $p$, $E\{r_c|p'\} > 0$ for all $p'\neq 1$, and $E\{r_c|p'\} = 0$ at $p'=1$. Instead, if $\pi=1$, then $Er_{nc}=E\{r_c|p'\} = 0$. Therefore, given that $\mu(q^a)<q$, $\pi=0$.[10] Similarly, if $\mu(q^{na})<q$, then $\pi=1$.[11]

Note that there are no equilibria other than the above. This can be shown by contradiction. Suppose that $\pi=1$ and $\mu(q^{na})=q$. Since $y\leq 1$, at $p'=0$ $p'(1-y)+yr^a < 1-p'$ (recall that by Lemma 2 and (2), $r^a=\frac{1}{2}$ when $\pi=1$),

---

[10] Note that at $p'=1$ the competent incumbent is therefore indifferent between $a$ and $na$. But choosing $a$ only at $p'=1$ violates Lemma 2.

[11] To verify that the pair: $(0,\mu(q^a))$, where $\mu(q^a)<q$, is rational consistent, we need to check that, by Bayes' rule, $q^a<q$ when $a$ is chosen with a small probability (instead of zero). Specifically, let $\pi^\varepsilon = (1-q)\pi^\varepsilon_{nc} + q \int_{p'=0}^{1} f(p')\pi^\varepsilon_c(p')dp' = \varepsilon$, where $\varepsilon$ is a small positive number. Then, by Bayes' rule, $q^a = \frac{q \int_{p'=0}^{1} f(p')\pi^\varepsilon_c(p')dp'}{\varepsilon}$, which implies that, $q^a < q \leftrightarrow \pi^\varepsilon_{nc} > \int_{p'=0}^{1} f(p')\pi^\varepsilon_c(p')dp'$. Therefore, $(0,\mu(q^a))$ such that $\mu(q^a)<q$, is rational consistent given the inequality above. Note that for any $\pi^\varepsilon_{nc}$ and $\pi^\varepsilon_c(p')$ that satisfies this inequality, $\pi^\varepsilon_{nc} \to 0$, and $\pi^\varepsilon_c(p') \to 0$ for all $p'$, as $\varepsilon\to 0$. Similarly this can be verified for the pair: $(1,\mu(q^{na}))$, where $\mu(q^{na})<q$.

and, therefore, by (4), the competent incumbent deviates to $\pi_c(0)\neq 1$, a contradiction.

Now suppose that $\pi=0$ and $\mu(q^a)=q$. Then, by (4), $E\{r_c|1\} = 0$. However, if the competent incumbent deviates to $\pi_c(1)=1$, then, by Lemma 1, $r^a=1$ and therefore, by (4), $E\{r_c|1\} = 1$, a contradiction. QED

Proof of Proposition 3: If $p \leq \frac{1-y}{2-y}$, then:

$$\int_{p'=0}^{\frac{1-y}{2-y}} f(p')p'\,dp' + \int_{p'=\frac{1-y}{2-y}}^{1} f(p')p\,dp'$$

$$< \int_{p'=0}^{\frac{1-y}{2-y}} f(p')p'\,dp' + \int_{p'=\frac{1-y}{2-y}}^{1} f(p')\,p'dp'$$

$$= p.$$

If $p > \frac{1-y}{2-y}$, then:

$$\int_{p'=0}^{\frac{1-y}{2-y}} f(p')p'\,dp' < \int_{p'=0}^{\frac{1-y}{2-y}} f(p')\,pdp'$$

$\leftrightarrow$

$$\int_{p'=0}^{\frac{1-y}{2-y}} f(p')p'\,dp' + \int_{p'=\frac{1-y}{2-y}}^{1} f(p')p\,dp' < p. \text{ QED}$$

Proof of Proposition 4: Consider a backward looking voter. By Lemma 1, the strategies played in Proposition 1 ensures that $p^{el}>p$ when $S$ is not revealed, and therefore $r_{bl}=1$ when $S$ is not revealed and $\underline{p}\leq p$. Given that $r_{bl}=1$ when $S$ is not revealed, by the proof of Proposition 1, these strategies are optimal.

Consider a forward looking voter. His equilibrium behavior is given by Proposition 2, since the proof of Proposition 2 apply regardless of whether $p \gtreqless \underline{p}$. QED

**Appendix 2: Discussion of a backward looking voter for whom $p>\underline{p}$:**

In this case, equilibrium may or may not differ from the one characterized in Proposition 1. To see this, suppose that $\underline{p}>p$ and the incumbent plays the strategy in Proposition 1. If $p^{el}>\underline{p}$ when $S$ is not revealed, then the same reasoning used in the proof of Proposition 4 establishes the equilibrium of proposition 1 as an equilibrium for this case as well. If $p^{el}<\underline{p}$ when $S$ is not revealed, then $r^a=0$. If $r^a=0$, then by (4) and (5), the optimal strategy of the incumbent is to use the threshold $1/(2-y)$ (instead of the threshold $(1-y)/(2-y)$ in Proposition 1). This threshold strategy is then an equilibrium iff it remains true that $p^{el}<\underline{p}$, which depends on the parameter $p$ and on the probability distribution $f(p')$ (see (6)).

As in the case of a pro-action voter, for either one of the equilibria above, comparing welfare is not possible without further assumptions.

**References**

Achen, C.H., Bartels, L.M., 2004. Blind retrospection: Electoral responses to

    drought, flu, and shark attacks.

Ashworth, S. and De Mesquita, E.B., 2014. Is voter competence good for voters?: Information, rationality, and democratic performance. *American Political Science Review*, pp.565-587.

Battaglini, M. and Harstad, B., 2020. The political economy of weak treaties. *Journal of Political Economy*, *128(2)*.

Caselli, F., Morelli, M., 2004. Bad politicians. *Journal of Public Economics*, *88*(3-4), pp.759-782.

Cohen, C. and Werker, E.D., 2008. The Political Economy of ``Natural'' Disasters. *Journal of Conflict Resolution*, 52(6), pp.795-819.

Dal Bó, E., Finan, F., Folke, O., Persson, T., Rickne, J., 2017. Who becomes a politician?. *The Quarterly Journal of Economics*, *132*(4), pp.1877-1914.

Galasso, V., Nannicini, T., 2011. Competing on good politicians. *American Political Science Review*, *105*(1), pp.79-99.

Gasper, J.T., Reeves, A., 2011. Make it rain? Retrospection and the attentive electorate in the context of natural disasters. *American Journal of Political Science*, *55*(2), pp.340-355.

Healy, A., Malhotra, N., 2009. Myopic voters and natural disaster policy. *American Political Science Review*, *103*(3), pp.387-406.

Herrera, H., Ordoñez, G. and Trebesch, C., 2020. Political booms, financial

crises. *Journal of Political Economy*, *128*(2).

Kramer, G.H., 1971. Short-term fluctuations in US voting behavior, 1896–1964. *American Political Science Review*, *65*(1), pp.131-143.

Levy, G. and Razin, R., 2015. Correlation neglect, voting behavior, and information aggregation. *American Economic Review*, *105*(4), pp.1634-45.

Malhotra, N. and Kuo, A.G., 2008. Attributing blame: The public's response to Hurricane Katrina. *The Journal of Politics*, 70(1), pp.120-135

Maskin, E., Tirole, J., 2004. The politician and the judge: Accountability in government. *American Economic Review*, *94*(4), pp.1034-1054.

Miller, R.K., Field, C.B. and Mach, K.J., 2020. Barriers and enablers for prescribed burns for wildfire management in California. *Nature Sustainability*, pp.1-9.

Millner, A., Ollivier, H. and Simon, L., 2020. Confirmation bias and signaling in Downsian elections. *Journal of Public Economics*, *185*, p.104175.

Scharfstein, D.S. and Stein, J.C., 1990. Herd behavior and investment. *The American Economic Review*, pp.465-479.

Wolfers, J., 2002. Are voters rational?: Evidence from gubernatorial elections. Graduate School of Business, Stanford University.
25